\documentclass[12pt]{article}  
\newcommand{\be}{ \begin{equation}}
\newcommand{\ee}{\end{equation}} 
\begin{document} 
\def\theequation{\arabic{section}.\arabic{equation}} 
\begin{titlepage} 
\title{A common misconception about {\em LIGO} detectors of 
gravitational waves} 
\author{Valerio Faraoni\\ \\ 
{\small \it  Physics Department, Bishop's University}\\
{\small \it 2600 College Street, Sherbrooke, Quebec, Canada 
J1M~0C8}\\
{\small \it Electronic mail: vfaraoni@ubishops.ca}\\
}
\date{} \maketitle 
\vspace*{1truecm} 
\begin{abstract} 
A common misconception about laser interferometric detectors of 
gravitational waves purports that, because the wavelength of 
laser light and the length of an interferometer's arm are both 
stretched by a gravitational wave, no effect should be 
visible, invoking an analogy with cosmological redshift in an 
expanding universe. The 
issue is clarified with the help of a direct calculation.
\end{abstract} 
\vspace*{1truecm} 
\noindent 
\begin{center}  Keywords: gravitational waves
\end{center}
\end{titlepage}

\def\theequation{\arabic{section}.\arabic{equation}}


\section{Introduction}
\setcounter{equation}{0}
\setcounter{page}{2}

One of the most significant advances in experimental 
gravitation 
has been the development  of giant laser interferometers for the  
detection of gravitational waves such as the {\em LIGO} 
\cite{LIGO}  
and {\em VIRGO} \cite{VIRGO} projects. The {\em LISA} 
project \cite{RowanHough} is even 
more  ambitious and plans the construction of a similar 
interferometer 
in space, with arms of size $\sim 5\cdot 10^6$~km,  an 
interplanetary 
distance. The recent coming on line of the {\em LIGO} detectors 
with actual data runs has brought the laser interferometric 
detectors to the forefront of current gravitational research. 
There is an objection as to how an interferometric detector  
works that recurs often, and persists as a misconception, among   
physics students and professionals alike. The 
objection is:  ``Given 
that the gravitational field stretches both 
the interferometer  arm and the wavelength of laser light 
propagating along it, why is the effect of a gravitational 
wave detectable? After all, the same situation occurs in 
cosmology, when the expansion of space stretches all distances 
and the wavelength of light alike, causing cosmological 
redshift''.

The immediate  answer to this objection is that the calculation 
of the phase shift $\Delta \phi $ between the laser beams of  
a laser interferometer produces a result  that is 
gauge-independent, while the interpretation in terms of 
stretching of an interferometer's arm and of the wavelength of 
light depends on the gauge adopted, and only gauge-independent 
results are acceptable in physics. However, this is truly an 
indirect answer and it may be preferable  to provide a direct 
argument in the same gauge (TT gauge) used for the 
above-mentioned interpretation of the workings of {\em LIGO}.

We could  find three instances in the literature in which 
this issue is raised:\\
\noindent 1) A paper by Saulson \cite{Saulson} addresses the 
issue and provides a qualitative intuitive explanation.\\
\noindent 2) In a Cal Tech lecture course on gravitational waves 
available on the internet \cite{caltech}, K.S. Thorne addresses 
the issue 
reporting that this 
is the most common question asked about laser interferometers, 
and he provides a qualitative answer: ``Does the wavelength of 
the light in the gravitational wave get stretched and 
squeezed the same manner as these mirrors move back and 
forth? ... The answer is no, the spacetime curvature 
influences the light in a different manner that it influences 
the mirror separations ... the influence on the light is 
negligible and it is only the mirrors that get moved back and 
forth and the light's wavelength does not get changed at all 
...''. However,  substantiating Thorne's answer 
with  a clear mathematical argument is not entirely trivial, as 
is  shown in Sec.~3.\\
3) In a recent paper \cite{garfinkle} the issue is raised 
again, together with the 
purported  analogy with  the cosmological situation. The author  
discusses an analogy between the gauge freedom of 
general relativity and the Aharonov-Bohm effect in quantum 
mechanics. 
The message is that in 
both situations gauge-dependent quantities appear in the 
equations describing the physics but the final physical results 
calculated are gauge-independent. This is again the answer that 
at the end the only physical quantity measured (the 
phase shift $\Delta \phi$ between two laser beams in an 
interferometric detector) is gauge independent.

Saulson's pedagogical paper \cite{Saulson} is clear and very 
physical but  the argument presented is qualitative --- the 
``stretching  at a different rate'' of the light wavelength and 
of an interferometer's arm can be shown explicitly.  On 
the other hand, the argument proposed in Ref.~\cite{garfinkle} 
requires the use of the  Aharonov-Bohm  effect foreign to 
classical relativity and truly  unnecessary to understand laser 
interferometers. The 
answer provided is indirect and does not specifically address 
the question of ``why there is a net effect if the 
wavelength of light and the interferometer's arm are 
both stretched?''  It would be more gratifying if a direct 
argument were provided showing how the (proper) 
length of an interferometer's arm and the (proper) wavelength of 
laser  light are ``stretched at different rates'' by a 
gravitational  wave, which is what we set out to do in this 
paper.


\section{Laser interferometers}
\setcounter{equation}{0}

Before we proceed to discuss a reply to the objection 
stated above, we summarize in this section 
the basics of laser interferometric detectors of 
gravitational waves \cite{MTW,RowanHough}, which we need 
for reference in the following.

Consider a laser interferometer consisting of two perpendicular 
arms aligned along the $x$ and $y$ axis, respectively. In the 
absence of gravitational waves the two arms have exactly equal 
lengths $L$. A beam splitter is located at  the origin $x=y=0$ 
and two mirrors are placed at the opposite end of each arm, at 
$x=L$ and $y=L$, respectively.  A monochromatic laser beam 
passing through the 
beam splitter is divided  into two beams that propagate along 
the $x$ and $y$-arms, are reflected by the mirrors, and travel 
back to the beam splitter where they are collected and 
compared, thus detecting any phase shift that may have occurred 
during the travel. If a gravitational wave impinges on the 
interferometer, it will cause  a phase shift $\Delta \phi$ 
between the two beams.

The gravitational wave is described as a 
small perturbation of the Minkowski metric $\eta_{\mu\nu}$. The 
spacetime metric is
\be \label{1}
g_{\mu\nu}=\eta_{\mu\nu}+h_{\mu\nu} \;,
\ee
where $\eta_{\mu\nu}=\mbox{diag}\left( -1,1,1,1 \right)$ and 
$\left| h_{\mu\nu} \right|\ll 1$ in an asymptotically Cartesian 
coordinate system. The transverse-traceless (TT) gauge is most 
often used in introductions to laser interferometric 
detectors. In this gauge $h_{0\mu}= {h^{\mu}}_{\mu}=0$ 
and the unperturbed metric is used to raise and lower indices. 
The only nonzero components of $h_{\mu\nu}$ in this gauge are 
$h_{xx}=-h_{yy}$ and $ h_{xy}=h_{yx}$, corresponding to two 
independent polarizations of the gravitational wave.  Only 
first order quantities in the metric perturbations $h_{\mu\nu}$ 
and their derivatives are considered because of the smallness of 
these quantities in any physical situation of interest. For 
simplicity we consider a gravitational wave with a single 
polarization traveling along the $z$-axis perpendicular to the 
interferometer's arms, perfectly reflecting mirrors, and a single 
reflection of each laser beam.  The geodesic deviation equation 
rules the evolution of the proper distances $x^i$ along the $x$ 
and $y$-axis: 
\be \label{2}
\ddot{x}^i={R^i}_{00j}\, x^j \;,
\ee
where $i,j=1,2$ and ${R^{\mu}}_{\nu\alpha\beta}$ is the Riemann 
tensor, which is most conveniently calculated in the TT gauge 
yielding \cite{MTW}
\be  \label{3}
\delta \ddot{x}^i= \frac{1}{2} \, \ddot{h}_{ij}^{(TT)}\, x^j \;.
\ee
A further  assumption almost always used in the literature on 
ground-based interferometric detectors is 
that the  wavelength $\lambda_{gw}$ of the gravitational wave  
is much  larger than the size of the interferometer, 
$\lambda_{gw} \gg L$, thus simplifying the integration of 
eq.~(\ref{3}) to
\begin{eqnarray}
\delta x &=& \frac{h_{xx}}{2}\, x \;,  \label{4}\\
&&\nonumber \\
\delta y &=& \frac{h_{yy}}{2}\, y \;. \label{5}
\end{eqnarray}
The difference in the variation of the proper lengths of the 
interferometer's arms  when a gravitational wave impinges along 
the $z$-axis with polarization $h_{xx}=-h_{yy}=h_{+}(t)$ is, to 
first order,
\be \label{6}
\delta l(t)=\delta x(t)-\delta y(t) =Lh_{+}(t)
\ee
and the phase difference between the two beams collected at the 
origin is 
\be \label{7}
\Delta \phi=2\pi \, \frac{\delta l}{\lambda} =2\pi \, 
\frac{L}{\lambda} h_{+}(t) \;,
\ee
where $\lambda$ is the wavelength of the monochromatic laser 
light.

\section{Variation of the wavelength of laser light in the 
$x$-arm}
\setcounter{equation}{0}

In this section we compute explicitly the variation of the 
wavelength of laser light propagating along one arm (say, the 
$x$-arm) of the laser 
interferometer, and then we compare the result with the 
variation of 
the proper length of this arm.  It turns out that the two 
quantities are different and this result shows that the 
objection ``the 
wavelength and the arm are stretched by the same amount as in 
the expansion of the universe'' is really unfounded. The TT 
gauge commonly used to 
calculate and interpret the effect of the gravitational wave is 
employed.

The variation of the wavelength of monochromatic laser light in 
the $x$-arm of the interferometer can be derived by considering 
the equation of null geodesics
\be \label{8}
\frac{dk^{\mu}}{d\tau} 
+\Gamma^{\mu}_{\rho\sigma}k^{\rho}k^{\sigma}=0 
\;,
\ee
where $k^{\mu}= dx^{\mu}/d\tau $ is the geometric tangent to the 
null geodesic and $\tau $ is an 
affine parameter along it. The  photon four-momentum is 
$p^{\mu}=\omega k^{\mu}= \left( \omega, \vec{k} \right)$,  where
$\omega $ and $\vec{k}$ are the angular frequency 
and the three-dimensional wave vector, respectively.  The 
Christoffel symbols are 
\be \label{9}
\Gamma^{\mu}_{\rho\sigma}=
\frac{1}{2} \, \eta^{\rho\alpha} \left( 
h_{\alpha\rho , \sigma}+
h_{\alpha\sigma , \rho}
- h_{\rho  \sigma , \alpha} \right) +\mbox{O}(h^2) \;.
\ee
The unperturbed laser beam 
travels along the $x$-axis with four-tangent  
$ k_{(0)}^{\mu}=  \delta^{0\mu}+\delta^{1\mu} $ 
while the actual (perturbed) four-tangent is $ k^{\mu}=
k_{(0)}^{\mu} +\delta k^{\mu} $ with $ \delta 
k^{\mu}=\mbox{O}(h)$. To first order, the deflections $
 \delta k^{\mu} $ satisfy the equation
\be \label{11}
\frac{d ( \delta 
k^{\mu})}{d\tau}=-\frac{1}{2} \, \eta^{\mu\alpha}\left( 
h_{\alpha \rho ,\sigma}+
h_{\alpha \sigma ,\rho} -
h_{\sigma \rho ,\alpha} \right) k_{(0)}^{\rho} k_{(0)}^{\sigma} 
\;,
\ee
where the product 
\be \label{12}
k^{\rho} k^{\sigma} =k_{(0)}^{\rho} k_{(0)}^{\sigma} 
+\mbox{O}(h)=
\left( \delta^{0\rho}\delta^{0\sigma}+
\delta^{0\rho}\delta^{1\sigma}+
\delta^{1\rho}\delta^{0\sigma}+
\delta^{1\rho}\delta^{1\sigma} \right) +\mbox{O}(h) 
\ee
can be taken to zero order --- including first order corrections 
only contributes second order terms to the deflections. 
Integration of eq.~(\ref{11}) with respect to $x$ between the 
beam splitter ($x=0$) and the  mirror ($x=L$) along the 
interferometer's arm yields the total deflection between the 
injection of the beam and its reflection at the mirror
\begin{eqnarray} 
\delta k^{\mu}&=& - \int_0^L dx\left( {h^{\mu}}_{\rho , 
\sigma} -\frac{1}{2} {h_{\rho\sigma}}^{,\mu} \right)
\left( \delta^{0\rho}\delta^{0\sigma}+
\delta^{0\rho}\delta^{1\sigma}+
\delta^{1\rho}\delta^{0\sigma}+
\delta^{1\rho}\delta^{1\sigma} \right) +\mbox{O}(h^2) \nonumber 
\\
&& \nonumber \\
&=&
- \int_0^L dx\left( {h^{\mu}}_{0 , 0} 
+ {h^{\mu}}_{1 , 0}
+ {h^{\mu}}_{0, 1}
+ {h^{\mu}}_{1 , 1} \right) \nonumber \\
&+&  \frac{1}{2}\int_0^L dx 
\left( h_{0 0}+2 h_{0 1}+h_{11} \right)^{, \mu} +\mbox{O}(h^2) 
\;.  \label{13}
\end{eqnarray}
In eq.~(\ref{13}) the integration with respect to the affine 
parameter $\tau $ has been replaced by an integration with 
respect to $x$, introducing only a second 
order error.  
In TT gauge $ h_{0\mu}=0$ and therefore the first three terms 
in the first integral vanish. The remaining term 
$ {h^{\mu}}_{1,1}$  contributes a boundary term
$h_{11}\left( x=0 \right)-h_{11}\left( x=L \right) $. Since 
gravitational waves of wavelength $\lambda_{gw} \gg L$ are 
usually considered in laser interferometers the spatial 
variation of the gravitational wave in the interferometer's arms 
is negligible and this term vanishes as well.  In TT gauge we 
are left with 
\be \label{14}
\delta k^{\mu}=\frac{1}{2}\int_0^L dx \, 
{h_{11}}^{,\mu}+\mbox{O}(h^2) \;.
\ee
The deflection of light traveling between the beam splitter at 
$x=0$ and the detector  at the same place (after 
reflection  of the signal at $x=L$) is given by
\be \label{15}
\delta k^{\mu}=\frac{1}{2}\int_0^{2L} dt \, 
{ h_{11} }^{,\mu}+\mbox{O}(h^2) 
\ee
by taking into account the fact that for the unperturbed laser 
beam traveling along the $x$-arm it is $x=t$ between emission 
and reflection, and $x=L-t$ between reflection and detection 
(the presence of the gravitational wave introduces corrections 
that only give second order terms in the deflections). Under 
the assumption $\lambda_{gw} \gg L$ the spatial dependence of 
$h_{\alpha\beta}$ can be neglected and $h_{\alpha\beta}\left( t, 
\vec{x} \right) \simeq h_{\alpha\beta}(t)$, therefore
\be \label{16}
\delta k^{\mu}=\frac{ \delta^{0\mu}}{2} \left[ 
h_{11}\left( t=2L \right)- 
h_{11}\left( t=0 \right) \right] +\mbox{O}(h^2) \;.
\ee
Therefore, in the approximation used, the laser photons do not 
suffer spatial deflections to first order. This fact is usually 
tacitly assumed in the standard presentations of how laser 
interferometers work (see \cite{polariz} for a 
discussion) but it appears explicitly in the approach used here.

In general, the angular frequency of the light measured by an 
observer with four-velocity $u^{\mu}$ is 
$\omega=-p^{\mu}u_{\mu}$. This quantity is a scalar and 
therefore is gauge-invariant. Let $u^{\mu}_0$ be the 
four-velocity 
in the absence of gravitational waves and $u^{\mu}=u^{\mu}_{(0)} 
+\delta u^{\mu}$ the perturbed four-velocity. Then 
\be \label{17}
\omega=-g_{\mu\nu} p^{\mu}u^{\nu}=-
\eta_{\mu\nu} p^{\mu}_{(0)} u^{\nu}_{(0)}-\eta_{\mu\nu}\left(
p^{\mu}_{(0)} \delta u^{\nu}+
 u^{\mu}_{(0)} \delta p^{\nu}  \right) -
h_{\mu\nu} p^{\mu}_{(0)} u^{\nu}_{(0)}
+\mbox{O}(h^2) \;.
\ee
In particular, by taking as the ``observer'' the 
half-transparent mirror at the origin, the unperturbed 
four-velocity is $u^{\mu}_{(0)}=\delta^{0\mu}$ and $\delta 
u^{\alpha}=0$.  In fact, in TT gauge the coordinate 
locations of freely falling bodies are 
unaffected by the gravitational wave, although their {\em 
proper} separations do vary. Then, upon detection at the beam 
splitter, the angular frequency of the laser light is
\be \label{18}
\omega=\omega_0 -\eta_{\mu\nu}\delta^{0\nu} \delta p^{\nu} 
-\omega_0 h_{\mu\nu} \left( \delta^{0\mu} \delta^{0\nu}
+\delta^{1\mu} \delta^{1\nu} \right)+\mbox{O}(h^2)= 
\omega_0+\delta p^0 +\mbox{O}(h^2) 
\ee
and the percent frequency shift, which is gauge-invariant, is
\be \label{19}
\frac{\delta\omega}{\omega_0}=\frac{ 
\omega-\omega_0}{\omega_0}=\frac{ \delta 
p^0}{\omega_0}=\frac{ h_{11}\left( t=0 \right)-
h_{11}\left( t=2L \right) }{2} \;.
\ee
In the approximation $\lambda_{gw}\gg L$ the temporal variation 
of $ h_{11}\left( t\right) $ during the short time $ \approx 2L$ 
it takes  for the light to travel to the mirror and back is 
negligible and $\delta \lambda \simeq 0$ in this approximation. 
Now we want to compare the variation of the  proper length of the 
interferometer arm with the variation in wavelength, and the way 
to do this correctly is to compare the former with the percent 
variation in the {\em proper} (or physical) wavelength of light 
$\lambda_{phys}\equiv \sqrt{ g_{11}}\, \lambda$. This is 
analogous to what is normally done in an expanding  
Friedmann-Lemaitre-Robertson-Walker spacetime with curvature 
index $K=\pm 1$ or $0$, described by the 
line element
\be
ds^2= -dt^2+a^2(t)\left[ \frac{dr^2}{1-Kr^2} +r^2 \left( 
d\theta^2+\sin^2\theta \, d\varphi^2 \right) \right] 
\ee
in comoving coordinates $\left( t,r, \theta, 
\varphi \right) $. In this case one can consider the wavelength 
of light $\lambda$ 
in comoving coordinates but what is physically relevant is the 
comoving (or physical) wavelength $\lambda_{phys}\equiv a(t) 
\lambda$ which changes with the comoving time $t$.  Similarly, 
in the interferometer's 
arm, the physical distance between mirror and beam splitter is 
not the coordinate distance $L$ (which is constant in TT 
gauge) but rather $\sqrt{g_{11}}\, L= \left[ 1+\frac{h_{11}}{2} 
+\mbox{O}(h^2) \right] L $ and the physical wavelength is 
$\lambda_{phys}=\sqrt{g_{11}} \, \lambda$. 
The percent variation of proper wavelength however 
coincides with the percent variation of coordinate wavelength,
\be
\frac{ \delta \lambda_{phys}}{ \lambda_{phys}}=
\frac{ \sqrt{g_{11}} \, \delta \lambda }{\sqrt{g_{11}} \,  
\lambda}= \frac{ \delta \lambda }{ \lambda} 
\ee
and therefore is also zero to first order in the approximation 
$\lambda_{gw} \gg L$.


\section{Discussion}
\setcounter{equation}{0}

The percent variation of the proper length of the 
interferometer's $x$-arm is given by eq.~(\ref{4}) as 
\be \label{21}
\frac{ \delta x(t)}{L} =\frac{1}{2} \, h_{+}(t) \;.
\ee
In the approximation $\lambda_{gw}\gg L$  the time dependence 
disappears  and
\be \label{22}
\frac{ \delta x}{L} =\frac{ h_{+}(t=0)}{2} \;,
\ee
which is different from zero, while the percent variation of 
proper wavelength $\delta \lambda_{phys}/\lambda_{phys} $ for 
the laser 
light  traveling in this arm is zero in this approximation.

Therefore the objection that ``all 
lengths are stretched at the same rate by the gravitational 
wave'' and based on the analogy with the expanding three-space 
of cosmology, is incorrect. The gravitational wave 
``treats in a different way''  the wavelength of light and the 
length of the interferometer's arm. Physically, the 
interferometer works by measuring the differential stretching 
of the $x$ and $y$ arms while the high frequency  light wave 
essentially experiences no inhomogeneities in the ``medium'' in 
which it propagates --- the gravitational wave --- because the 
wavelength  $\lambda_{gw}$ of the gravitational wave is 
so much larger than the wavelength of light. This 
conclusion agrees with Thorne's qualitative answer to the 
objection \cite{caltech}. Technological issues aside, laser 
interferometers such as those of {\em LIGO} and {\em VIRGO} can 
indeed  detect  gravitational waves.

\section*{Acknowledgments}

It is  a pleasure to thank C.J. Pritchet and D.A. Hartwick for 
stimulating discussions. This work is supported by the Natural 
Sciences  and  Engineering Research  Council of Canada (NSERC).

 
\end{document}